\documentclass[11pt]{article}
\usepackage{jheppub}
\usepackage{url,comment}
\usepackage{times}
\usepackage{latexsym}
\usepackage{graphicx, graphics, amsmath, amssymb, slashed, xcolor, bbm,bm,amsthm, array}
 \usepackage{subfigure}
 \usepackage{listings} 
 \usepackage{trimclip}
\usepackage{rotating}
\usepackage{afterpage}
 
\usepackage{multirow} 

\usepackage{hyperref}	
\hypersetup{  colorlinks=true,     linkcolor=blue,     filecolor=magenta,          urlcolor=blue}

\newcommand{\nc}{\newcommand}

\nc{\mc}{\mathcal}
\nc{\er}[1]{(\ref{eq:#1})}
\nc{\onehalf}{\frac{1}{2}} 
\nc{\partialbar}{\bar{\partial}}
\nc{\psit}{\widetilde{\psi}}
\nc{\Tr}{\mbox{Tr}}
\nc{\hc}{\mbox{H.c.}}
\nc{\ev}{\;\mathrm{eV}}
\nc{\mev}{\;\mathrm{MeV}}
\nc{\gev}{\;\mathrm{GeV}}
\nc{\kev}{\;\mathrm{keV}}
\nc{\tev}{\;\mathrm{TeV}}
\nc{\pev}{\;\mathrm{PeV}}
\nc{\eev}{\;\mathrm{EeV}}

\def\chii0{\chi_i^0}
\def\chij0{\chi_j^0}

\newcommand{\gsim}{\lower.7ex\hbox{$\;\stackrel{\textstyle>}{\sim}\;$}}
\newcommand{\lsim}{\lower.7ex\hbox{$\;\stackrel{\textstyle<}{\sim}\;$}}
\nc{\ttbar}{t\bar t}

\newcommand{\cref}[1]{Chapter~\ref{c.#1}}

\graphicspath{{plots/}}

\title{
Recent Progress and Next Steps for the MATHUSLA LLP Detector
}

\author{
\includegraphics[width=4cm]{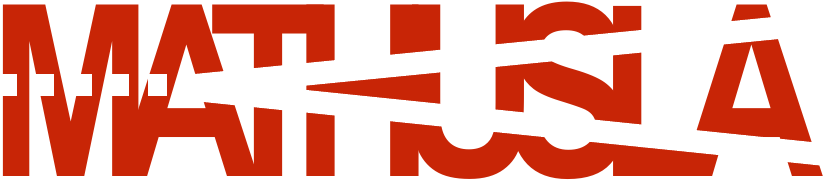}
\\
\vspace{5mm}
{\normalfont 
Snowmass 2021 Letter of Interest (EF09, EF10, IF9)
\\
\vspace{5mm}}}

\author[1]{\hspace*{-1mm}\textnormal{Authors:}\\Cristiano Alpigiani,}
\author[2]{Juan Carlos   Arteaga-Vel\'azquez,}
\author[3]{Austin   Ball,}
\author[4]{Liron   Barak,}
\author[5]{Jared   Barron,}
\author[6]{Brian   Batell,}
\author[7]{James   Beacham,}
\author[4]{Yan   Benhammou,}
\author[45]{Benjamin Brau}
% ...

\affiliation[1]{University of Washington, Seattle, USA}
\affiliation[2]{Universidad Michoacana de San Nicol\'as de Hidalgo, Mexico (UMSNH)}
\affiliation[3]{CERN, Switzerland}
\affiliation[4]{Tel Aviv University, Israel}
\affiliation[5]{University of Toronto, Canada}
\affiliation[6]{University of Pittsburgh, USA}
\affiliation[7]{Ohio State University, USA}
% ...
\affiliation[45]{University of Massachusetts Amherst, USA}
% ...

\emailAdd{mathusla.experiment@cern.ch}

\abstract{

We report on recent progress and next steps in the  design of the proposed MATHUSLA Long Lived Particle (LLP) detector for the HL-LHC as part of the Snowmass 2021 process. Our understanding of backgrounds has greatly improved, aided by detailed simulation studies, and significant R\&D has been performed on designing the scintillator detectors and understanding their performance. The collaboration is on track to complete a Technical Design Report, and there are many opportunities for interested new members to contribute towards the goal of designing and constructing MATHUSLA in time for HL-LHC collisions, which would increase the sensitivity to a large variety of highly motivated LLP signals by orders of magnitude. 

}

\begin{document}

\begin{flushright}
\phantom{\small{.}}
\end{flushright}

\maketitle

\section{Introduction}
\label{s.introduction}

MATHUSLA (Massive Timing Hodoscope for Ultra-Stable neutraL pArticles)~\cite{Chou:2016lxi, Alpigiani:2018fgd, Alpigiani:2020fgd} is a proposed large-scale dedicated Long-Lived Particle (LLP) detector to be situated at CERN near the CMS detector. It will be able to reconstruct the decay of neutral LLPs, produced in HL-LHC collisions in CMS, as displaced vertices (DV) in a near-zero-background environment. The physics case for LLP searches at the HL-LHC in general and MATHUSLA in particular was explored in detail in~\cite{Curtin:2018mvb}. MATHUSLA would be able to extend the sensitivity in long lifetime and LLP cross section by several orders of magnitude compared to the main LHC detectors alone, depending on the production and decay mode.
In particular, it would allow searches for LLPs with lifetimes near the upper Big Bang Nucleosynthesis bound set by cosmology, and also play a vital role in the search for Dark Matter (DM). 
MATHUSLA is therefore crucial in exploiting the full physics potential provided by existing LHC collisions.

In this Snowmass contribution we summarize the current status of the MATHUSLA proposal and recent developments made by the MATHUSLA experimental collaborations on the detector design, associated R\&D, and detailed simulations that are in progress to understand backgrounds and their suppression strategies. We also briefly review the LLP physics reach, as well as MATHUSLA's surprising ability to characterize any newly discovered physics in great detail if it can act as a L1 trigger for CMS.

\section{Overview of MATHUSLA Detector}
\label{s.overview}
\begin{figure}[h!]
    \begin{center}
        \includegraphics[width=0.9\textwidth]{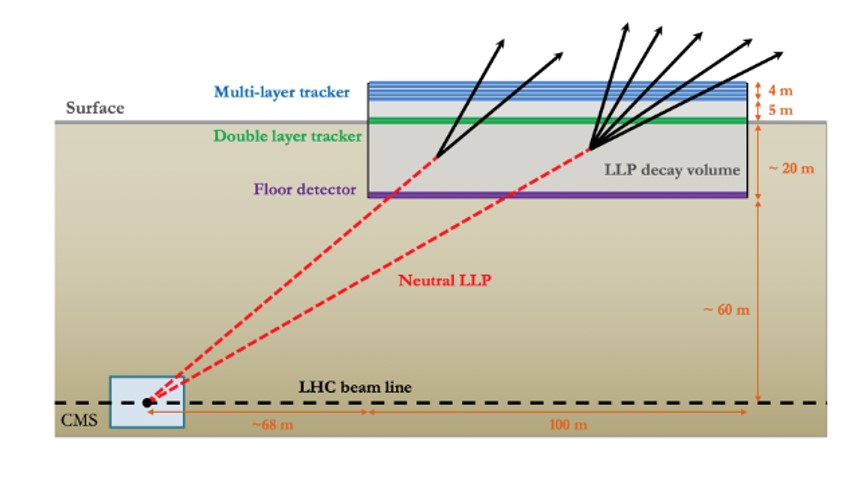}
        \caption{Illustration of the concept of MATHUSLA}
   \label{fig:cartoon}
   \end{center}
\end{figure}

\begin{figure}[h!]
    \begin{center}
        \includegraphics[width=0.9\textwidth]{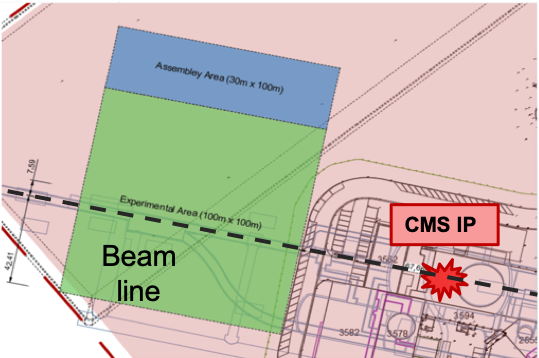}
        \caption{MATHUSLA hall (green) at LHC P5}
   \label{fig:layout_P5}
   \end{center}
\end{figure}

Figure \ref{fig:cartoon} shows an illustration of the concept of MATHUSLA. It is a very large detector sited on the surface near CMS. Having a large part of the decay volume underground brings it closer to the IP, which increases the solid angle in the acceptance for LLPs generated in the collisions. Figure \ref{fig:layout_P5} shows the proposed layout of the experimental hall on the LHC P5 site, close to CMS. The 100 m $\times$ 100 m experimental area and the 30 m $\times$ 100 m adjacent area for the detector assembly is shown in green and blue, respectively. The structure, which is located on the surface near the CMS IP fits well on CERN owned land. To adjust to the available land, this proposal has a 7.5 m offset to the center of the beams. The site allows for the detector to be as close as 68 m from the IP, which is shown in red. 

\begin{figure}[h!]
    \begin{center}
        \includegraphics[width=0.9\textwidth]{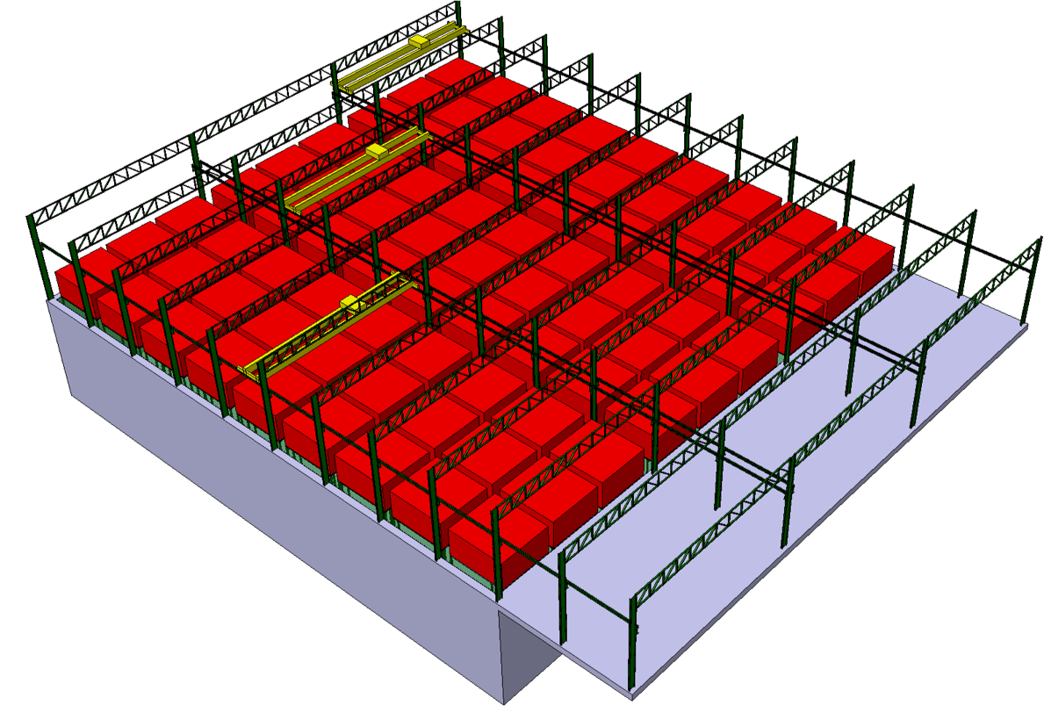}
        \caption{Overview of MATHUSLA. In the MATHUSLA experiment there will be 81 identical modules in an array. Each module is 9 m $\times$ 9 m $\times$ 30 m on an approximate 10 meter spacing.The experiment will be installed in a surface building at LHC P5 near the CMS experiment. The experimental hall will be excavated 20 m below surface and extend about 10 meters above ground.}
   \label{fig:Overview}
   \end{center}
\end{figure}

\begin{figure}[h!]
\centering 
\includegraphics[width=.3\textwidth]{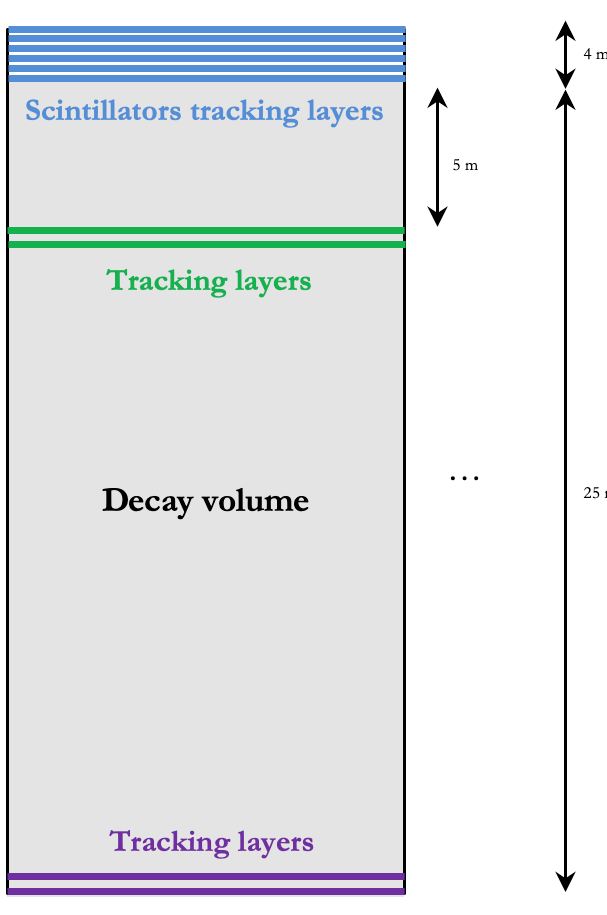}
\qquad
\includegraphics[width=.6\textwidth]{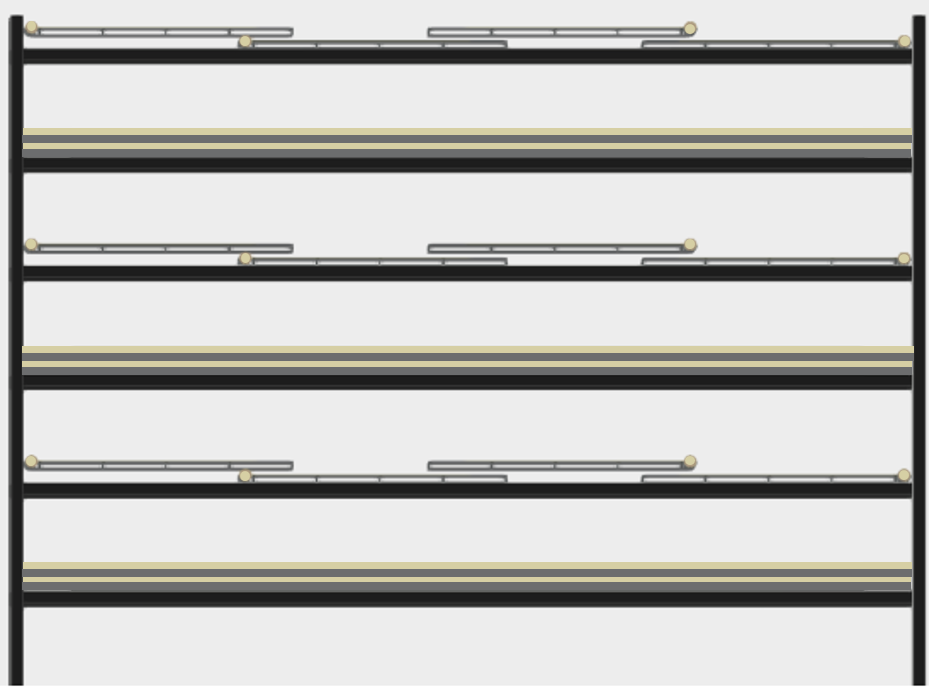}
\caption{ Left: One of the 81 identical MATHUSLA modules. There are 10 layers of scintillating planes.  Right: Detail of the top 6 scintillating planes in a module. Alternate layers have the extrusions running at 90 degrees. Note that each layer of scintillating planes is composed of 16 separate extrusion planes. A small overlap between the extrusion planes guarantees full detection coverage. }
\label{fig:module design}
\end{figure}

Figure~\ref{fig:Overview} shows the overview of the MATHUSLA experiment and a possible layout inside its experimental hall. It will be a 100m $\times$ 100m $\times$ 30m active volume, to be sited near the CMS detector at the LHC. MATHUSLA will be composed of 81 identical modules, approximately 9 m $\times$ 9 m $\times$ 30 m each. Figure~\ref{fig:module design} shows details of a module. There will be 10 layers of scintillators in the module, composed of pieces of extruded scintillator bars. We are considering scintillating bars that are 1 cm thick, about 250 cm length and width in the range 3.5 cm to 5 cm. Pairs of extrusions will be read out by a wavelength-shifting (WLS) fiber coupled to two SiPMs, as is done for instance in Mu2e. 

\begin{figure}[h!]
    \begin{center}
        \includegraphics[width=0.8\textwidth]{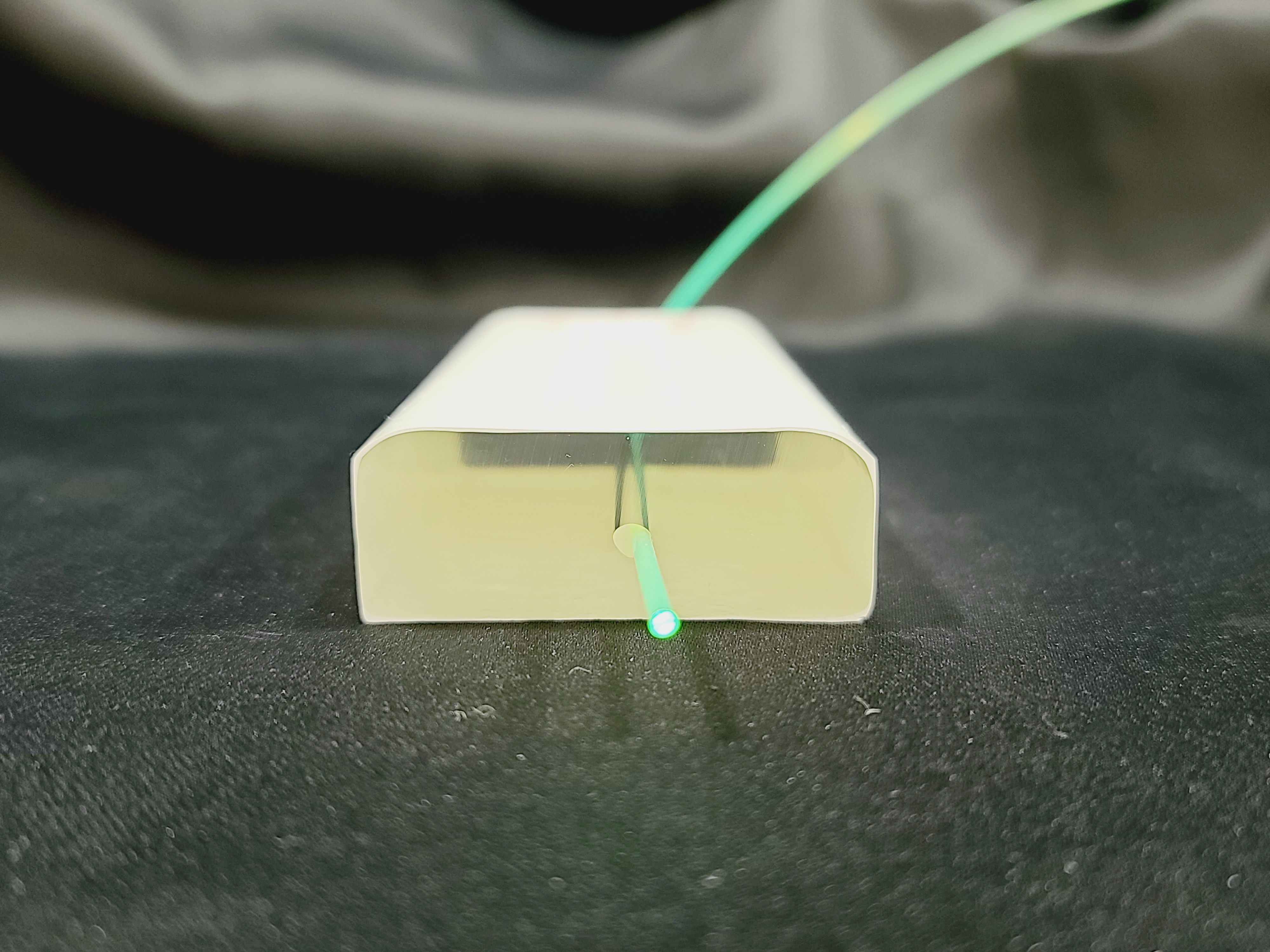}
        \caption{An extrusion with one central fiber running through a co-extruded hole. Note the co-extruded white reflective coating surrounding the bar.}
   \label{fig:Extrusion}
   \end{center}
\end{figure}

\begin{figure}[h!]
    \begin{center}
        \includegraphics[width=0.9\textwidth]{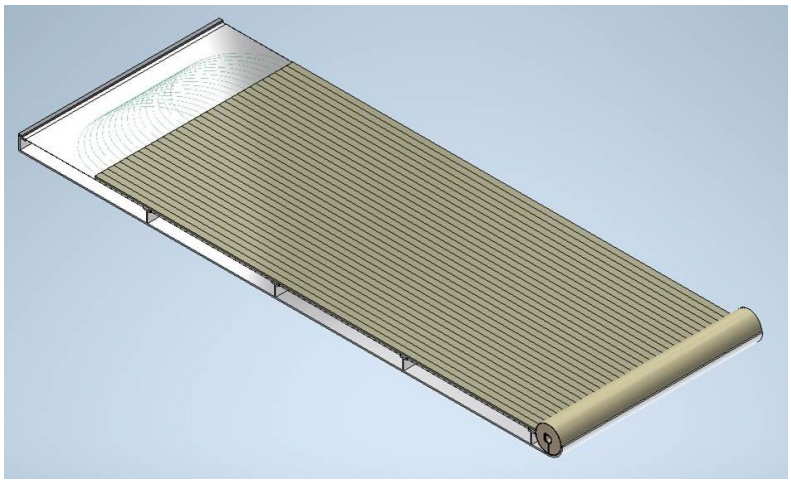}
        \caption{One of the units making up one of the four sub-planes in a scintillator layer.}
   \label{fig:Extrusion_plane}
   \end{center}
\end{figure}

\begin{figure}[h!]
    \begin{center}
        \includegraphics[width=0.9\textwidth]{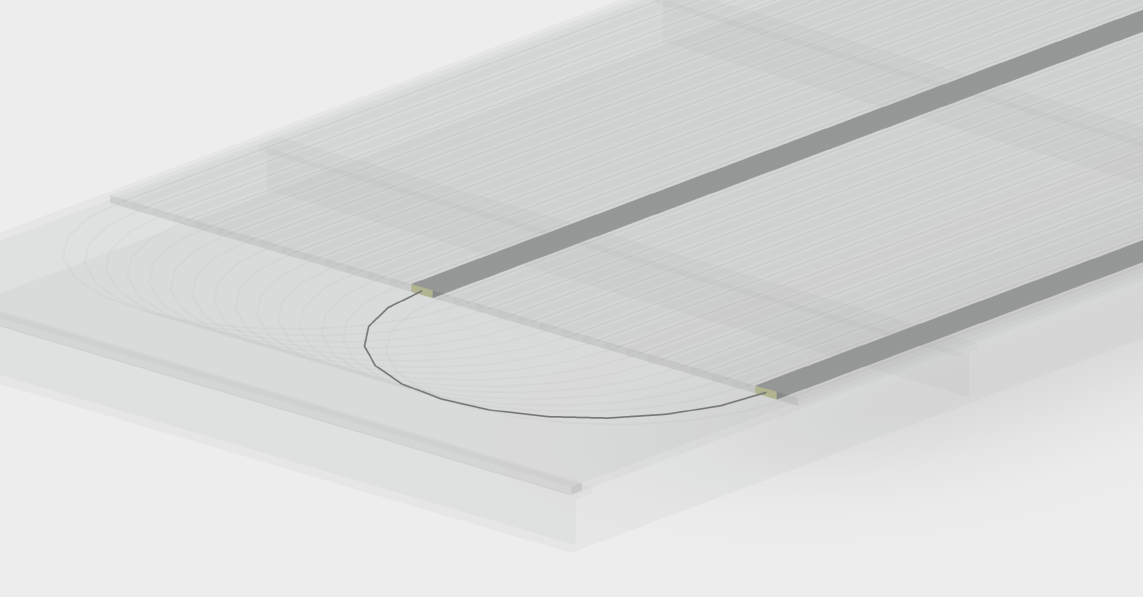}
        \caption{This is the interior end of the scintillation extrusion plane. Fibers make a 180 degree bend and return in a different extrusion. The separation of extrusions is determined to satisfy the minimum bend radius of the fiber. This design results with all active components (SIPMs, electronics) at accessible ends of the scintillation layer. Additionally this design eases the trigger formation as both ends of the fiber are physically closer to each other.}
   \label{fig:fiber_bend}
   \end{center}
\end{figure}

Figure~\ref{fig:Extrusion} shows a representative extrusion with fiber inserted into a co-extruded hole running along the middle of the extrusion. Figure~\ref{fig:module design} right side shows details of the scintillating layers in the module. Each scintillating layer is made of four sub-planes that tile the 9X9 m$^2$ area. These are about 2.5 meters long and 2.25 cm wide. They overlap slightly in the 2.5m direction to prevent cracks, as shown in Figure~\ref{fig:module design}, right hand side. The direction of the extrusions are rotated by 90 degrees for alternating scintillating layers that gives X-Y segmentation.  Measuring the time difference between light pulses at the ends of the WLS fiber provide complementary coordinate measurement in each scintillating layer. An extrusion plane is shown in Figure~\ref{fig:Extrusion_plane}. These planes are mechanically independent and are assembled and tested in workshops remote from CERN. During installation into the modules at CERN, cooling and electrical attachments are connected. In total there will be 16*10*81 = 12960 identical extrusion planes.
Note on one end of the extrusion plane is an area for the WLS fiber to bend and return in a different scintillating extrusion bar. In this way all SiPMs and electronics are on the same end of the extrusion plane.  Details of the fiber-bend region are shown in Figure~\ref{fig:fiber_bend}. The cylindrical region shown in Figure~\ref{fig:Extrusion_plane} contains the SIPMs, connections to electronics, and cooling for temperature stabilization. The extrusions and the electronics/cooling system are mounted on an aluminum structure to provide support. Each extrusion plane has SIPMs with associated bias voltage and readout. The complete MATHUSLA will have approximately 730K channels of readout.

\clearpage

\section{Backgrounds}
\label{s.backgrounds}

A variety of sources of backgrounds to LLP signatures exist in the MATHUSLA detector, both from proton-proton collisions at the IP as well as ambient sources such as cosmic ray showers. Timing with nanosecond precision and high-coverage floor detectors play critical roles in identifying backgrounds and separating them from potential new physics sources.

\subsection{Backgrounds from the Interaction Point}

Over the lifetime of the HL-LHC, $\sim6\times10^{10}$ W bosons will be produced at the CMS IP that subsequently decay into a high-energy muon, of which approximately 1\% will have sufficient momentum and the proper trajectory to reach the MATHUSLA detector. Of those IP muons that reach the detector, roughly half are from W$\rightarrow\mu\nu$ events, with the second largest contribution coming from sequential decays of $\mathrm{b}\bar{\mathrm{b}}$. High-energy muons can form a reconstructed vertex in the MATHUSLA detector primarily through the scattering of a secondary, knock-off electron (delta ray) that is able to travel through an appreciable number of layers before ranging out. Electron/positron pair-production through the conversion of an internal or external photon from bremsstrahlung is also an important background process. Although a few hundred rare decays of $\mu\rightarrow eee\nu\nu$ will be present in the detector, it will not pose as a background since these muons need to be boosted in order to be reconstructable; however, the vast majority of those rare decays will originate from slow or captured muons.  In total, as many as $\sim 10^7$ reconstructable vertices from these processes are present in the detector that will need to be suppressed.

For LLP signatures with many charged particles in the final state (such as a neutral particle decaying to $\mathrm{q}\bar{\mathrm{q}}$), this background poses no difficulty as it rarely induces more than even two reconstructed tracks. However, for a decay signature such as $\mu^+\mu^-$, more effort is needed to distinguish the processes. The key technique is to have a high-efficiency floor veto to identify tracks as they pass through. Such a floor cannot have large gaps because multiple scattering degrades the intrinsic tracking resolution and tight fiducial requirements will degrade the signal acceptance. Topological and track/vertex quality criteria also play an important role in suppressing backgrounds. Tracks from delta rays rarely have significant penetration power and are often stopped before passing through the upper-most layers.  The opening angles between the reconstructed tracks are typically quite small and consistent with only the lightest of LLP candidates $m\lesssim 1$~GeV. Ongoing efforts in our simulation suggest that these backgrounds can be reduced to $\mathcal{O}(1)$ over the lifetime of the experiment for even boosted LLPs that decay into two tracks while still maintaining a high signal efficiency.

\subsection{Cosmic Ray Backgrounds}

Cosmic rays can produce backgrounds for the experiment in a number of ways. Although $\sim 10^{14}$ cosmic ray muons will pass through the detector over its lifetime, their trajectory is downward going, and the timing resolution is sufficient to prohibit mis-reconstruction in the opposite direction, let alone two tracks to form a vertex. Nevertheless, upwards going tracks can still be produced primarily through back-scattering off of the rock beneath the detector as well as the sides. For single tracks, these backgrounds are not substantial: the vast majority do not have appreciable range and are suppressible with the same techniques that IP muons are. A more pernicious background is the production of strangeness-bearing SM LLPs, such as $K_{\rm L}$ and $\Lambda$. The MATHUSLA collaboration is studying the rates for these processes from cosmic muons and protons in Geant4~\cite{Agostinelli:2002hh} simulations using the energy and angular distribution computed by Parma~\cite{PARMA3, PARMA4, PARMAsite}. Approximately $\mathcal{O}(10^7)$ such particles will be produced in the rock beneath the detector but re-emerge inside the detector before decaying. Nevertheless, the vast majority will not be reconstructable by the detector, as they are preferentially produced with large boost in directions parallel to ground, while those produced pointing towards the tracking layers are disproportionately soft. Several handles are available to suppress even this background: neutral particle production is correlated with nearby activity in the floor, and topological consistency with production at the IP is also a significant constraint.

Cosmic rays also induce atmospheric neutrino backgrounds in the detector that can be reconstructed as an LLP through inelastic and quasi-inelastic collisions with material in the detector. The cross section and integration time are long enough that several hundred inelastic atmospheric neutrino collisions should take place inside of the detector. Nevertheless, like many of these backgrounds, the collision products are especially soft. Since protons are often one of the accompanying particles in the collision, a simple restriction on $\beta>0.8$ for tracks is sufficient to suppress this background. Finally topological consistency with production at the IP is enough to bring this background for low-track multiplicity LLP candidates down to effectively 0.


\section{New Physics Reach}
\label{s.reach}

\begin{figure}
    \centering
    \includegraphics[width=0.6\textwidth]{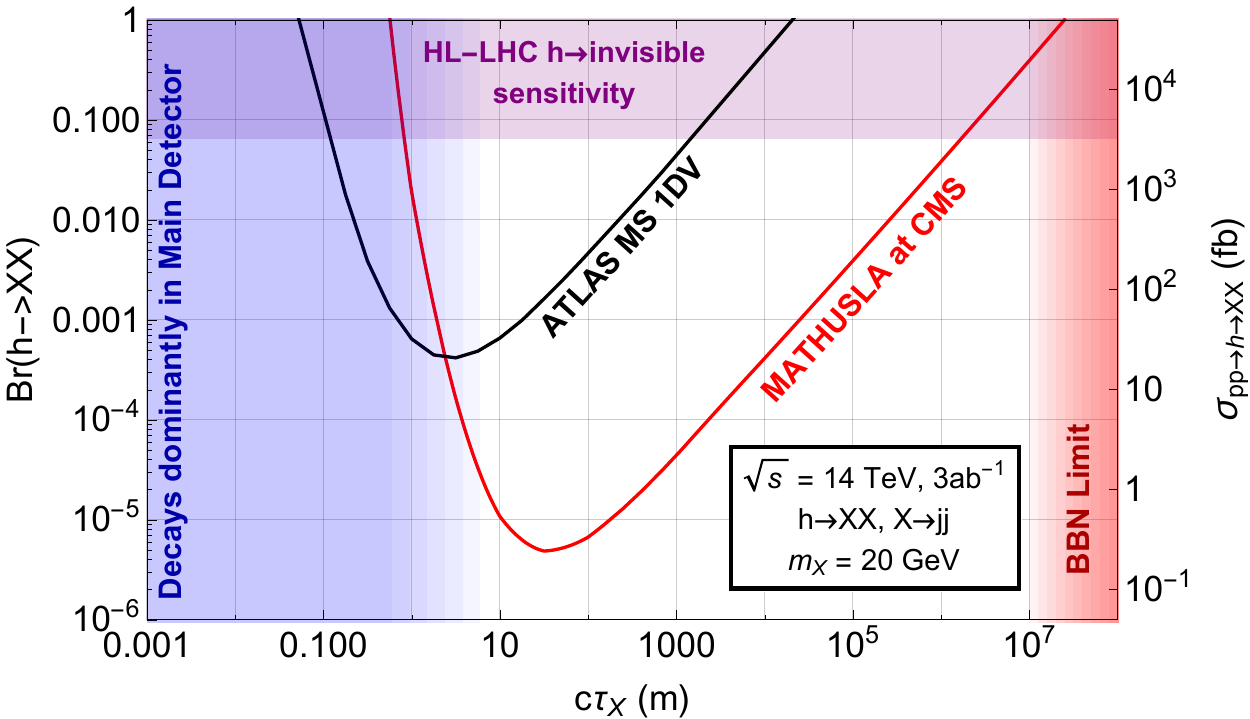}
    \caption{
    Red curve: MATHUSLA@CMS sensitivity (4 observed events) for LLPs of mass $m_X = 20 \gev$ produced in exotic Higgs decays. Black curve: reach of ATLAS search for a single hadronic LLP decay in the Muon System at the HL-LHC~\cite{Coccaro:2016lnz}.
    }
    \label{fig:sensitivity_higgs}
\end{figure}

\begin{figure}
    \centering
    \includegraphics[width=0.9\textwidth]{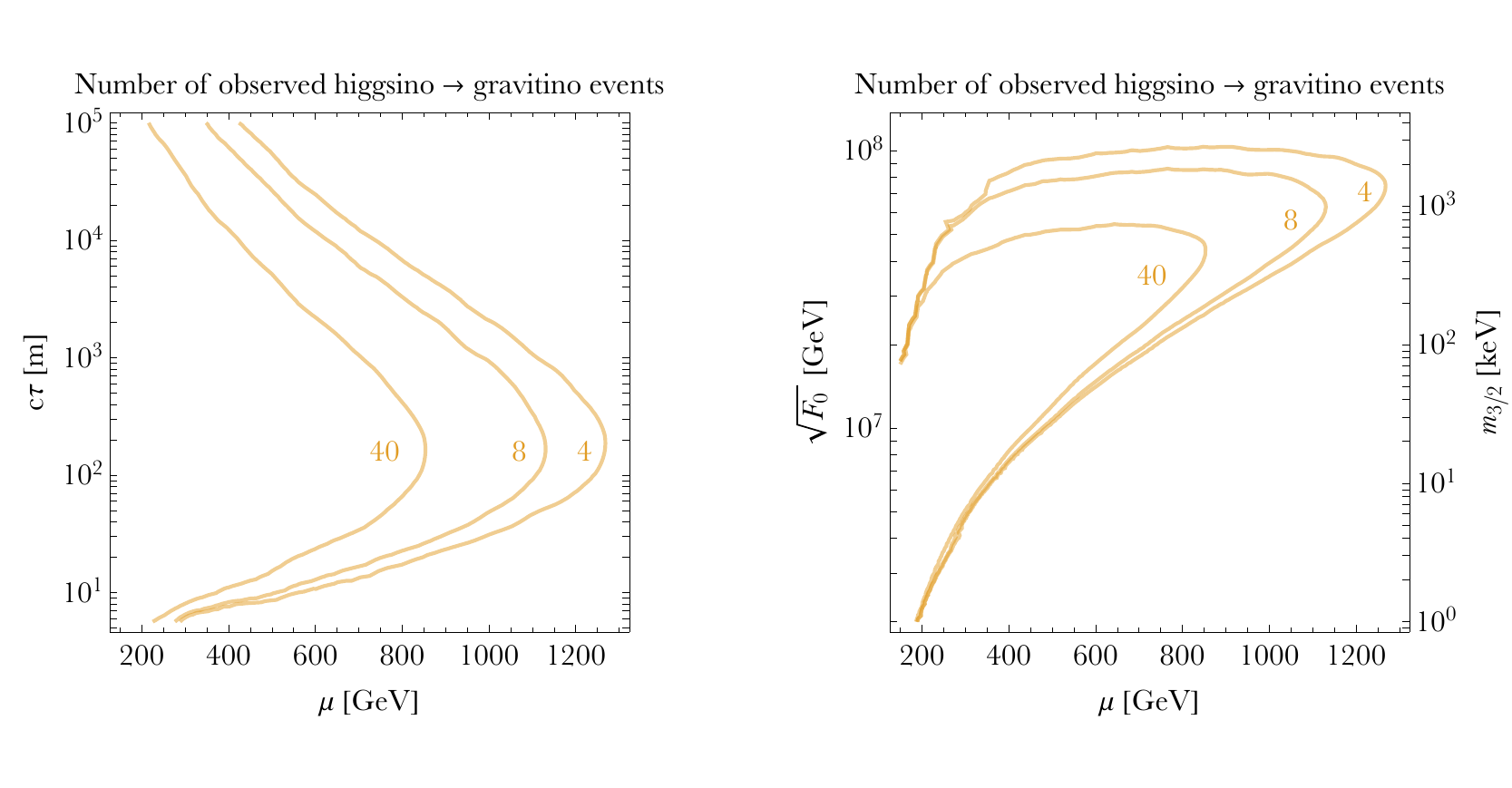}
    \vspace*{-5mm}
    \caption{
    Number of $\tilde H \to \tilde G + (Z,h)$ events that MATHUSLA@CMS could observe from electroweak production of higgsinos at the HL-LHC with an integrated luminosity of 3 ab$^{-1}$. \emph{Left:} higgsino mass $\mu$ versus lifetime $c \tau$ in meters. \emph{Right:} higgsino mass $\mu$ versus the SUSY breaking scale as parametrized by $\sqrt{F}$ in GeV (label on left axis) or gravitino mass $m_{3/2}$ in keV (label on right axis). In a wide swath of parameter space with higgsino lifetimes ranging from smaller than $10$~m to larger than $10^5$~m, MATHUSLA could provide a discovery of new physics with electroweak cross-sections for which the HL-LHC would fail to discover new physics. 
    }
    \label{fig:sensitivity_higgsinos}
\end{figure}

\begin{figure}
    \centering
        \hspace*{-23mm}
    \begin{tabular}{c}
     \includegraphics[width=0.6\textwidth]{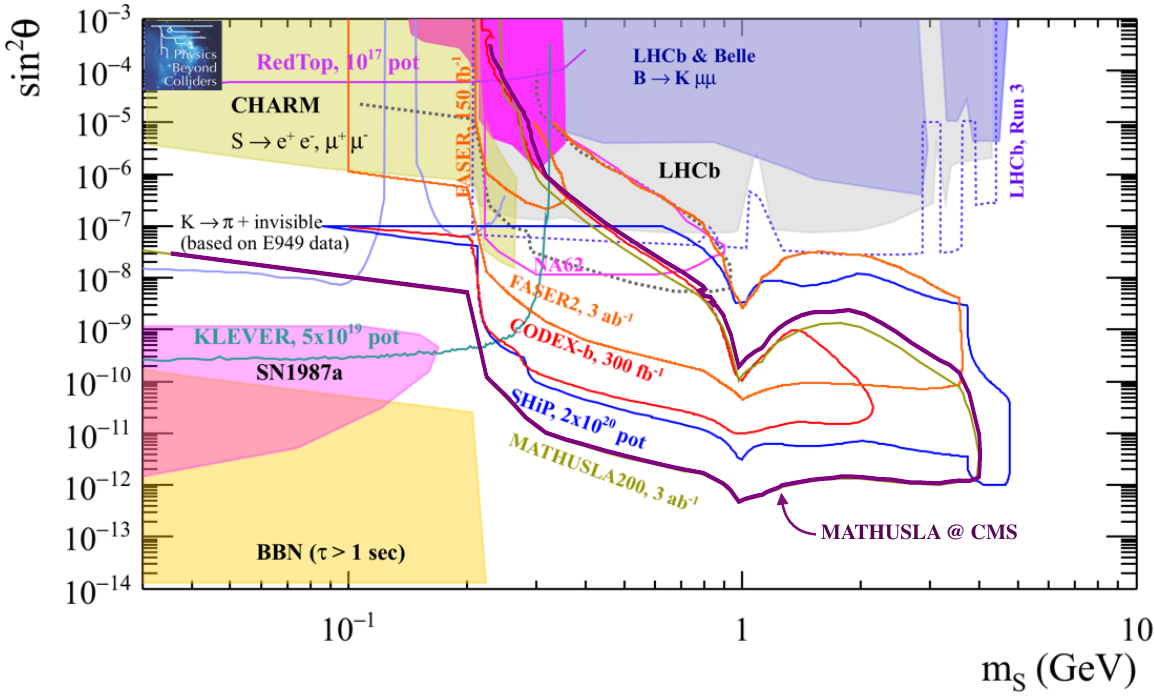}
     \vspace*{-5mm}
    \\
    (a)
    \vspace*{2mm}
    \\
    \begin{tabular}{cc}
      \includegraphics[width=0.6\textwidth]{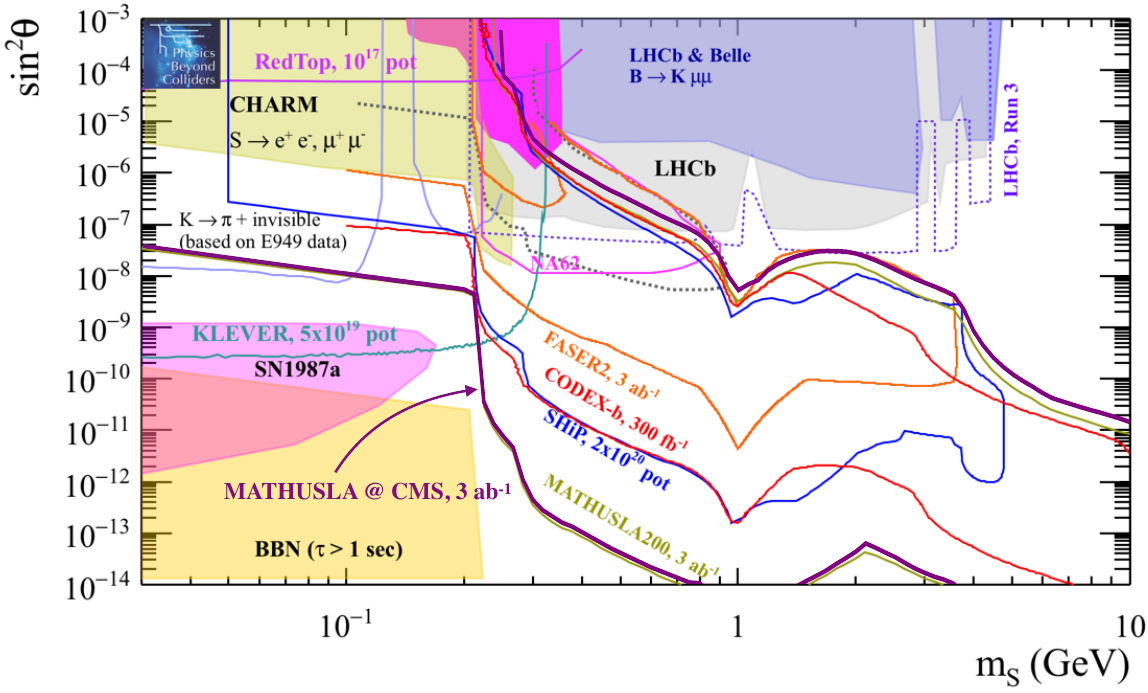}  
      &
        \includegraphics[width=0.6\textwidth]{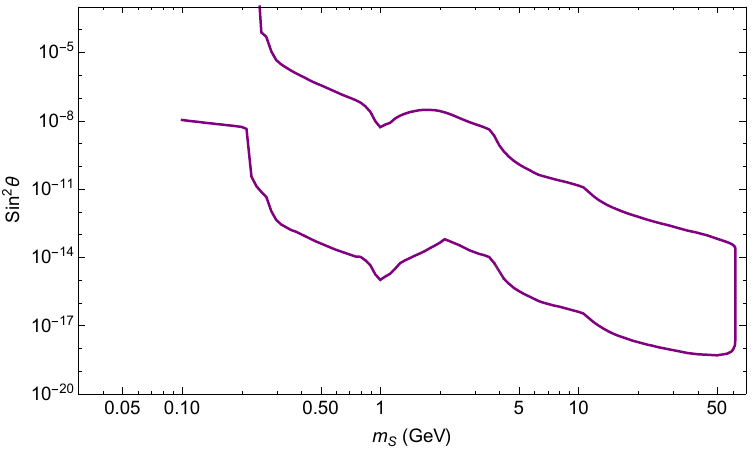}    \vspace*{-0mm}
    \\
    (b) & (c)
    \end{tabular}
    \end{tabular}
    \caption{
    Purple curves: sensitivity of MATHUSLA@CMS for a singlet scalar LLP $s$ mixing with Higgs mixing angle $\theta$. (a) Assuming production in exotic $B$, $D$, $K$ meson decays only. (b) Assuming additional production in exotic Higgs decays with $\mathrm{Br}(h\to ss) = 0.01$. Figures (a) and (b) are reproduced from the PBC BSM Working Group report~\cite{Beacham:2019nyx} with the purple MATHUSLA@CMS curves added. This  shows sensitivity of various other existing and proposed experiments, as well as the old MATHUSLA200 benchmark estimates (yellow curves). (c) Same scenario as (b) but showing the entire MATHUSLA sensitivity due to $h\to ss$ decays.}
    \label{fig:sensitivity_SMS}
\end{figure}

The motivation of LLP searches at the LHC in general and for constructing MATHUSLA in particular was extensively discussed in the physics case white paper~\cite{Curtin:2018mvb} and subsequent studies~\cite{Bauer:2018uxu,
Ibarra:2018xdl,
Nelson:2018iuc,
Curtin:2018ees,
Berlin:2018jbm,
Dercks:2018eua,
Ariga:2018uku,
Demidov:2018odn,
Beacham:2019nyx,
Wang:2019orr,
deNiverville:2019xsx,
Serra:2019omd,
Das:2019fee,
Boiarska:2019vid,
Chun:2019nwi,
No:2019gvl,
Krovi:2019hdl,
Bauer:2019vqk,
Wang:2019xvx,
Jana:2019tdm,
Hirsch:2020klk,
Banerjee:2020kww,
Kling:2020mch,
Barron:2020kfo,
Dreiner:2020qbi,
Cheng:2021kjg,
Bhattacherjee:2021rml,
Bertuzzo:2020rzo,
Du:2021cmt,
Archer-Smith:2021ntx,
Jodlowski:2021xye,
Hryczuk:2021qtz}. MATHUSLA can improve LHC reach by orders of magnitude in cross section and lifetime for LLP signal that are 
trigger- or background-limited at the main detectors.\footnote{For a recent discussion of LLP triggers at the LHC, see~\cite{Alimena:2021mdu}.}
In this section we briefly review sensitivity estimates for some of the most important LLP benchmark models to demonstrate MATHUSLA's reach for probing new physics by discovering LLPs produced at the full range of energies access by the LHC, from the TeV- to sub-GeV-scale.\footnote{The sensitivity estimates of the physics case white paper~\cite{Curtin:2018mvb} and many other studies assume the MATHUSLA geometry of the earliest proposals~\cite{Chou:2016lxi, Alpigiani:2018fgd}, which differs from current benchmark geometry targeted by the MATHUSLA collaboration. However, the LOI Update~\cite{Alpigiani:2020fgd} demonstrated that the LLP sensitivity of the current geometry is essentially identical, and the earlier sensitivity estimates can be applied verbatim.}

We first consider examples of weak- or TeV-scale LLPs produced at the LHC.
Figure~\ref{fig:sensitivity_higgs} shows the sensitivity to hadronically decaying LLPs produced in exotic Higgs decays, which arises in a large variety of new physics scenarios~\cite{Curtin:2013fra}, including solutions to the Hierarchy Problem like  Neutral Naturalness~\cite{Chacko:2005pe,Burdman:2006tz,Craig:2015pha,Curtin:2015fna}.
The LLP cross section sensitivity on the right axis approximately applies to most weak-scale LLP production processes~\cite{Curtin:2018mvb}. 
Figure~\ref{fig:sensitivity_higgsinos} shows the reach for meta-stable Higgsinos within Supersymmetry with gauge-mediated SUSY breaking~\cite{Giudice:1998bp,Knapen:2016exe}. 

A representative and minimal example of low-mass LLP reach is the singlet scalar LLP that has a tiny mixing angle $\theta$ with the Higgs boson, shown in Figure~\ref{fig:sensitivity_SMS}. (For details see~\cite{Evans:2017lvd,Beacham:2019nyx}). 
Part (b), where an exotic Higgs decay branching fraction of $\mathrm{Br}(h \to ss) = 0.01$ is assumed as an additional LLP production process, demonstrates the advantage gained by the high LHC energy even when searching for very light LLPs, since they can be produced in high-scale processes that are kinematically suppressed at intensity frontier experiments. 
Also shown are the expected sensitivities of other proposed experiments like FASER~\cite{Ariga:2018zuc}, CODEX-b~\cite{Aielli:2019ivi}, and SHiP~\cite{Alekhin:2015byh}.

%

We also emphasize that LLP searches are instrumental in the hunt for Dark Matter and are often the only way of observing the DM directly, as demonstrated by several recent studies~\cite{No:2019gvl,DAgnolo:2018wcn,Aielli:2019ivi,Berlin:2018jbm}.
In models like Freeze-In DM (FIDM)~\cite{Hall:2009bx, No:2019gvl}, inelastic DM (iDM)~\cite{TuckerSmith:2001hy, Izaguirre:2015zva}, co-annihilating DM~\cite{DAgnolo:2018wcn} or co-scattering DM~\cite{Aielli:2019ivi}, the relic abundance of the stable DM candidate is determined by the properties of an LLP in the thermal plasma of the Big Bang.
This LLP carries the same quantum number which stabilizes DM, and decays into DM + SM final states.
The DM particle itself could be almost completely sterile, precluding a direct detection signal, and production of the parent LLP at colliders could then be the only way to produce and observe DM.

Note that these sensitivity estimates assume perfect  detection efficiency as long as the LLP decays in the decay volume, and assume zero backgrounds after the rigorous geometric DV reconstruction cuts have been applied. 
As we discussed in the previous sectioon, the  zero-background assumption can be satisfied due to stringent geometric and timing requirements on the displaced vertex, as well as the use of various veto strategies. 
We have explicitly verified in simulations that the assumption of perfect reconstruction efficiency is a reasonable one for high-multiplicity final states of the LLP decay, which is the case for MATHUSLA's most important physics target, hadronically decaying LLPs with masses in the $\mathcal{O}(10 \gev) - \mathcal{O}(100 \gev)$ range. 
LLPs that decay leptonically or very light LLPs with masses $\lsim$ GeV that only decay to a small number of charged tracks may have somewhat lower reconstruction efficiency, but this can be ameliorated in various ways and is highly sensitive to the details of the final detector design. This issue is under active study in the MATHUSLA collaboration.

%
Finally, we point out that MATHUSLA is not only a new physics discovery machine, it can also measure the properties of newly discovered particles in great detail.
This was the subject of~\cite{Barron:2020kfo} (see also~\cite{Curtin:2017izq,Ibarra:2018xdl}), which demonstrate how MATHUSLA can characterize the new physics by working in tandem with the CMS main detector. This allows for determination of LLP decay mode, mass, production mode, and underlying parameters of the new physics theory like the parent particle mass with only $\mathcal{O}(10 - 100)$ observed LLP decays. 
Unlocking this surprisingly broad analysis capability, given that MATHUSLA does not collect energy or momentum information, requires MATHUSLA to act as a Level-1 Trigger for CMS, allowing for the correlation of information from both detectors.

\section{LLP Trigger and Data Acquisition}
\label{s.LLPtriger}

The Long-lived Particle trigger is based on two or more upward-going tracks that are consistent with a decay vertex in the MATHUSLA decay volume. 
There are two components in the trigger: a low-latency hardware-based Level-1 Trigger (L1T) and a software-based High Level Trigger (HLT). All L1T's for LLP are written to disk together with any other triggers such as those for monitoring and diagnostics and those for cosmic ray physics. All particle hits in the detector, without any pre-selection,  are similarly written to buffer storage. The HLT process subsequently defines a time window around each trigger, selects the corresponding particle hits, and saves them in archival storage for offline analysis.
The implementations for trigger and data acquisition (DAQ) are modular and scalable, and they are insensitive to layout changes of MATHUSLA detector modules. 

The Level-1 Trigger is divided into two steps: finding individual upward-going tracks and forming a vertex with these tracks. Track finding operates in parallel in each detector module, using hit information from the top six detector layers of that module as well as those from the eight surrounding modules making up a 3x3 array. A track candidate is defined by hits that are spatially aligned, and its direction of travel is given by relative hit times in different detector layers. Over the 5-m vertical extent of the trigger layers, upward-going signal tracks are separated from downward-going cosmic rays by ~30 nsec. They can be readily distinguished with the planned 1-nsec timing resolution. Upward-going tracks from all detector modules are passed to the vertex finder. A satisfactory 4-dimensional overlap of these tracks is then the Level-1 Trigger. Both track finding and vertex finding will be implemented in Field Programmatic Gate Array (FPGA) to satisfy the CMS trigger latency requirements in order to record the rest of the event recoiling against the LLP - see below.

Similar to track finding, each detector module in the MATHUSLA array writes out particle hits to buffer storage independently. 
Making the conservative assumption that fake hits is the same rate as the dominant true particle hit rate from cosmic rays, commercial off-the-shelf (COTS) disks are easily capable of the data volume and bandwidth requirements. For example, a 1-TB buffer storage disk is sufficient for up to one day's data volume. 

While the trigger and DAQ plans for MATHUSLA are generally straightforward and simple relative to other experiments, there is an interaction with CMS trigger that imposes a latency constraint on the MATHUSLA Level-1 Trigger. MATHUSLA by itself is fully capable of detecting Beyond Standard Model (BSM) physics exhibiting as long-lived particle decays. In order to go from observation of BSM physics to an understanding this BSM physics would benefit from information on the full 4-pi event that the CMS detector records. Thus, the MATHUSLA L1T must be available at CMS in time to participate in their trigger decision. Detailed estimates based on realistic firmware execution times, propagation delays, etc indicate that the latency time budget can be achieved. In particular, a prototype track finding algorithm is being implemented in FPGA and will be confirmed using simulated data.  

\clearpage
\section{Timing and Performance}
\label{s.Time resolution}

\begin{figure}[h!]
    \begin{center}
        \includegraphics[width=1.0\textwidth]{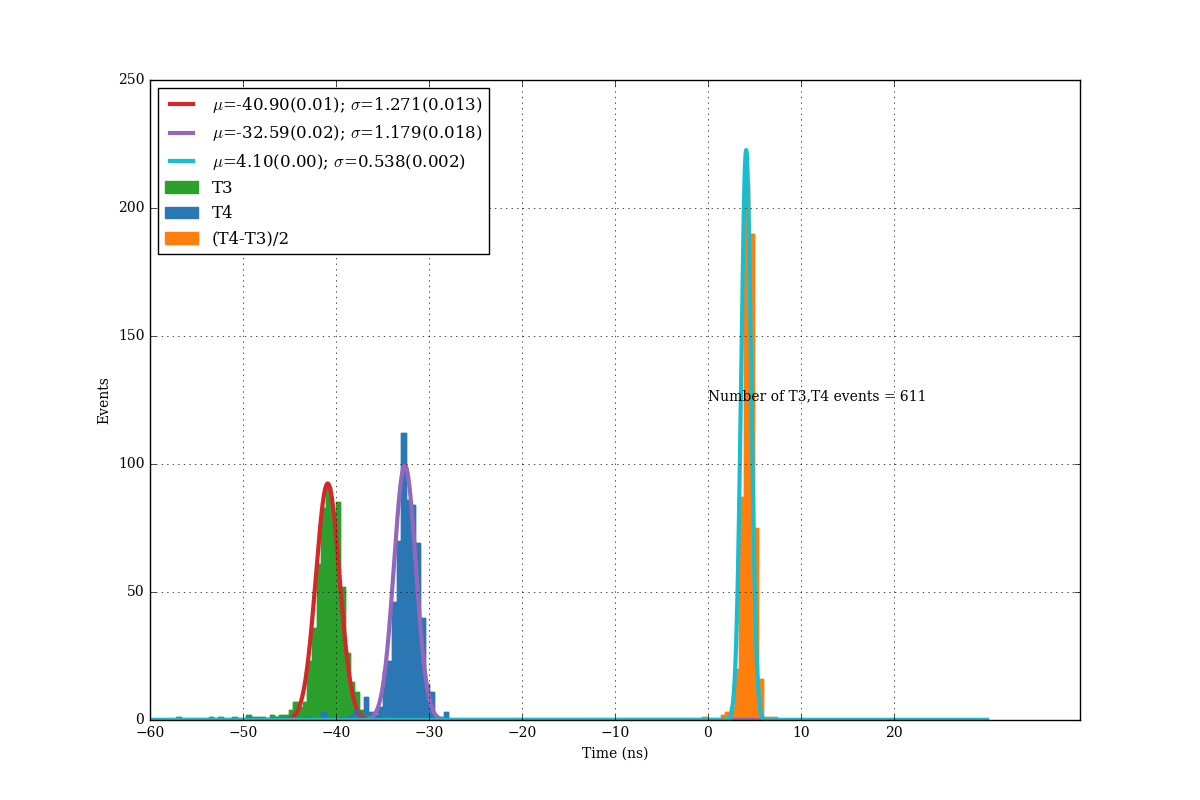}
        \caption{Timing measurement for a 5m long fiber through a 1X4cm extrusion. This location is at 250cm along the fiber, equidistant from the 2 SIPMs. Time distributions (relative to the cosmic trigger start time) are shown for the 2 SIPM channels (Chan 3, Chan 4). Also shown is the difference, (T4-T3)/2. We note this difference divided by 2 is our figure of merit for timing. The factor of 2 comes from the observation that different points along the fiber separated by delta have a +delta increase in distance from one SIPM, and a -delta decrease in distance from the other. The timing resolution of 0.538ns corresponds to about 9cm rms position resolution, well within MATHUSLA requirement.}
   \label{fig:timing}
   \end{center}
\end{figure}

MATHUSLA has 2 separate specifications for timing and for light yield. The timing performance is specified to achieve better than 15cm rms accuracy for cosmic rays hitting random positions along the scintillator bar.  We note that the speed of light in the fiber is about 17 cm$/$ns so timing resolution of less than 1ns for hit positioning is required. The light yield at the extreme case is required to be at least 15 photoelectrons per end of the fiber. This requirement is based on the need to be able to efficiently set an electronics threshold in the few PE range, to suppress accidental dark count triggers. The extreme case for light yield will be when the cosmic ray hits close to one end of the bar, with a short WLS fiber distance to the near SIPM, and a long WLS fiber distance to the far SIPM.
 The fiber is 5M long, representative of the length of fiber needed for the MATHUSLA detector. ( The fiber passes through a 2.5m extrusion and returns through a different extrusion in the same extrusion plane as illustrated in Figure~\ref{fig:fiber_bend}). Using cosmic ray studies we have confirmed a baseline design that satisfies our light yield and timing requirements. A full length 5m fiber was threaded through a representative section of 1X4 cm cross section extrusion and read out by SIPMs at each end.  Cosmic rays were triggered passing through the extrusion and various positions along the length of the fiber were studied. Shown in Figure~\ref{fig:timing} is a representative measurement of time resolution along a 5m fiber with the baseline 1X4cm extrusion cross section. This point is equidistant 250cm from both SIPMs. We have studied several WLS fiber types from different vendors and have found several that (combined with the extrusion size chosen) meet our timing and light yield requirements. The fiber in Figure~\ref{fig:timing} is a Kuraray 1.5mm diameter fiber. 1.5mm diameter fiber has been chosen for our baseline and the bend radius as shown in Figure~\ref{fig:fiber_bend} has been designed accordingly. The worst-case light yield for this configuration of extrusuion cross section and fiber was 23 PE, which greater than the requirement of 15PE.

\clearpage
\section{Scintillator Fabrication}
\label{s.scintunits}
\begin{figure}[h]
    \begin{center}
        \includegraphics[width=0.75\textwidth]{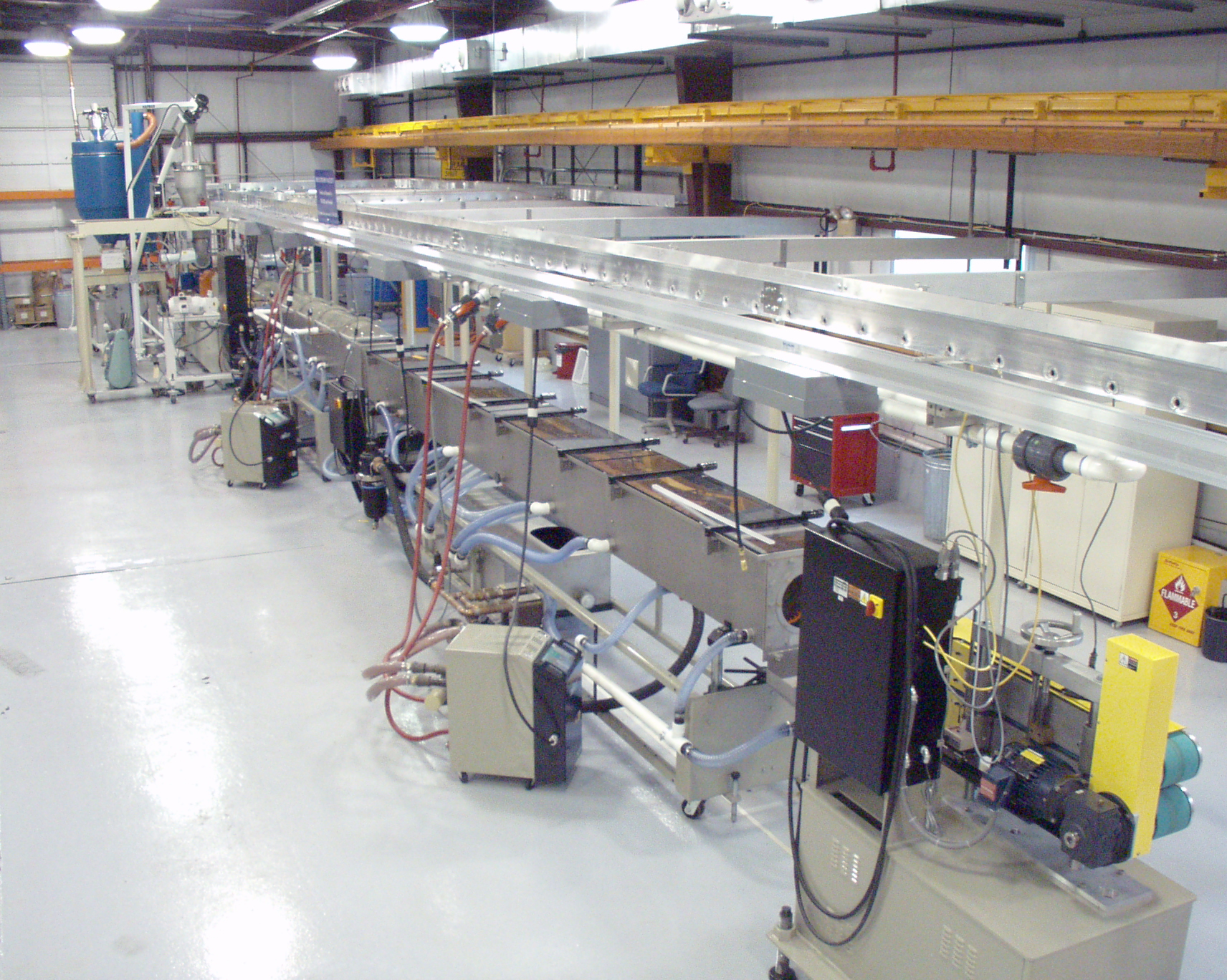}
        \caption{The scintillator extrusion facility at Fermilab.}
   \label{fig:Extrusion_line}
   \end{center}
\end{figure}

The scintillator extrusions will be fabricated at the NICADD facility at Fermilab. The scintillator extrusion system is shown in Figure~\ref{fig:Extrusion_line}. It is an in-line process where polystyrene pellets are mixed with the appropriate amount of fluorescent organic compounds and extruded through a die to create a scintillator extrusion of the profile required by the experiment. MATHUSLA requires about 700 tons of 1x4 cm profile scintillator extrusion. This corresponds to about 3 years of full production of the facility,

\clearpage


\bibliography{references}
\bibliographystyle{JHEP}

%

\end{document}